\documentclass[journal]{IEEEtran}
\usepackage{url}
\usepackage{graphicx}
\usepackage{amssymb}
\usepackage{color}
\usepackage{newalg}

\ifCLASSINFOpdf
\else
\fi
\begin{document}

%
\title{Cyber-Physical Security: A Game Theory Model of Humans Interacting over Control Systems}
%
%
%

\author{Scott~Backhaus,$^1$ Russell~Bent,$^1$ James~Bono,$^2$
        Ritchie~Lee,$^3$ Brendan Tracey,$^4$
        David~Wolpert,$^1$ Dongping~Xie,$^2$ and~Yildiray~Yildiz$^3$
\thanks{The authors are with $^1$Los Alamos National Laboratory, $^2$www.bayesoptimal.com, $^3$NASA Ames, and $^4$Stanford University.}
\thanks{Manuscript received XXXXX; revised XXXX.}}

%
%

\markboth{IEEE Transactions on Smart Grids,~Vol.~X, No.~X, XXXXX~XXXXX}%
{Backhaus \MakeLowercase{\textit{et al.}}: Cyber-Physical Security of Smart Grids}
%



\maketitle

\begin{abstract}
\boldmath
Recent years have seen increased interest in the design and deployment of smart grid devices and control algorithms. Each of these smart communicating devices represents a potential access point for an intruder spurring research into intruder prevention and detection. However, no security measures are complete, and intruding attackers will compromise smart grid devices leading to the attacker and the system operator interacting via the grid and its control systems. The outcome of these machine-mediated human-human interactions will depend on the design of the physical and control systems mediating the interactions. If these outcomes can be predicted via simulation, they can be used as a tool for designing attack-resilient grids and control systems. However, accurate predictions require good models of not just the physical and control systems, but also of the human decision making. In this manuscript, we present an approach to develop such tools, i.e. models of the decisions of the cyber-physical intruder who is attacking the systems and the system operator who is defending it, and demonstrate its usefulness for design.

\end{abstract}


%
\IEEEpeerreviewmaketitle

\section{Introduction}

\IEEEPARstart{S}upervisory Control and Data Acquisition (SCADA) systems form the cyber and communication components of electrical grids. Human operators use SCADA systems to receive data from and send control signals to grid devices to cause physical changes that benefit grid security and operation. If a SCADA system is compromised by a cyber attack, the attacker may alter these control signals with the intention of degrading operations or causing widespread damage to the physical infrastructure.

The increasing connection of SCADA to other cyber systems and the use of off-the-shelf computer systems for SCADA platforms is creating new vulnerabilities\cite{Cardenas_2008} increasing the likelihood that SCADA systems can and will be penetrated. However, even when a human attacker has gained some control over the physical components, the human operators (defenders) retain significant SCADA observation and control capability. The operators may be able to anticipate the attacker's moves and effectively use this remaining capability to  counter the attacker's moves. The design of the physical and control system may have a significant impact on the outcome of the SCADA operator's defense, however, designing attack resilient systems requires predictive models of these human-in-the-loop control systems. These machine-mediated, adversarial interaction between two humans have been described in  previous game-theoretic models of human-in-the-loop collision avoidance systems for aircraft\cite{TCAS} and our recent extensions of these models to electrical grid SCADA systems\cite{NIPS_Chapter}. The current work builds upon and extends this previous work.

The model of machine-mediated human-human interactions described in \cite{TCAS} includes two important components. The first is a mathematical framework for describing the physical state of the system and its evolution as well as the available information and its flow to both the humans and the automation. Well-suited to this task is a semi Bayes net\cite{TCAS} which, like a Bayes net, consists of: a set of {\it nodes} representing fixed conditional probability distributions over the physical state variables and the sets of information and {\it directed edges} describing the flow and transformation of information and the evolution of the physical state between the nodes. However, a semi Bayes net also includes ``decision'' nodes with unspecified conditional probability distributions that will be used to model the strategic thinking of the humans in the loop. When these decision nodes incorporate game-theoretic models, the resulting structure called a semi network-form game (SNFG) of human strategic behavior.

Game theoretic models of the humans are fundamentally different than models of the automation and control algorithms. These simpler devices process inputs to generate outputs without regard for how their outputs affect other components, nor do they try to infer the outputs of other components before generating their own output. Strategic humans perform both of these operations. In adversarial interactions, a strategically thinking human infers the decisions of his opponent and incorporates this information into his own decision making. He also incorporates that his opponent is engaged in the same reasoning. These behaviors distinguish humans from automation making the principled design of human-in-the-loop control systems challenging. In our model, we will utilize game theoretic solution concepts to resolve the circular player-opponent inference problem just described and compute the conditional probability distributions at the decision nodes in a SNFG representation of a SCADA system under attack.

A game theoretic model of a decision node includes two important components. The first is a utility or reward function that captures the goals of the human represented by the decision node and measures the relative benefit of different decisions. The second component is a solution concept that determines {\it how} the human goes about making decisions. As a model of human behavior, the solution concept must be selected to accurately represent the humans in question. For example, the humans may be modeled as fully rational, i.e. always selecting the action that maximizes their reward, or as bounded rational, sometimes taking actions that are less than optimal. Additionally, if the decisions that the humans are facing are too complex to afford exhaustive exploration of all options, the mathematical operations we use to represent the human's mental approximations are also part of the solution concept.

The current work builds upon previous game-theoretic models of human-in-the-loop aircraft collision avoidance systems\cite{TCAS} and our recent extension of  these models to simplified electrical grid SCADA systems\cite{NIPS_Chapter} where the focus was on developing the computational model for predicting the outcome of a SCADA attack where the SCADA operator was certain that an attacker was present. In the present work, we retain the simplified electrical grid model, but make several important extensions. First, we remove the SCADA operator's certainty that an attacker is present forcing the operator to perform well under both normal and ``attack'' conditions. Second, we shift our focus from only predicting the outcome of an attack to an initial effort at using these predictions as a tool to design physical and control systems. Third, the extension to design requires numerical evaluation of many more scenarios, and we have implemented new computational algorithms that speed our simulations.

To summarize, the {\it designer} models and simulates the behavior of the SCADA {\it operator} and the cyber-physical attacker by developing reward functions and solution concepts that closely represent the decision making processes of these humans. These game theoretic models are embedded into the decision nodes of a SNFG that represents the evolution of the physical state and information available to both human decision nodes and the automation nodes. If the model is accurate, then the {\it designer} can utilize this model to predict the outcomes of different system designs and, therefore, maximize his own ``designer's reward function". This design process closely resembles the economic theory of mechanism design \cite{Mas-Colell95, Osborne94}, whereby an external policy-maker seeks to design a game with specific equilibrium properties. The key difference between our work and mechanism design is that we do not assume equilibrium behavior, and this enables us to use the standard control techniques described above\cite{Wolpert08, Wolpert10, Bono11, Wolpert10b}.

We also make contributions to the growing literature on game theory and network security \cite{Manshaei13, Alpcan11}. The assumption that human operators infer the existence of an attacker from the state of the SCADA places this model alongside work on intrusion detection systems \cite{Zhu12, Liu05, Liu06, Lye02, Nguyen09}. However, we also model the human operator's attempts to mitigate damages when an attack is detected. So our model contributes to the literature on intrusion response \cite{Zonouz09}.

The rest of this paper is organized as follows. Section~\ref{sec:grid} describes the simplified electrical distribution circuit and the SCADA used to control it. Section III reviews the structure of SNFG and points out features and extensions important for the current work. Section IV describes the solution concept we apply to our SCADA model. Section V and VI describe the simulation results and our use of these results to assess design options, respectively. Section VII gives our conclusions and possible directions for future work.


\section{Simplified Electrical Grid Model}\label{sec:grid}
\begin{figure}
\includegraphics[scale=.5]{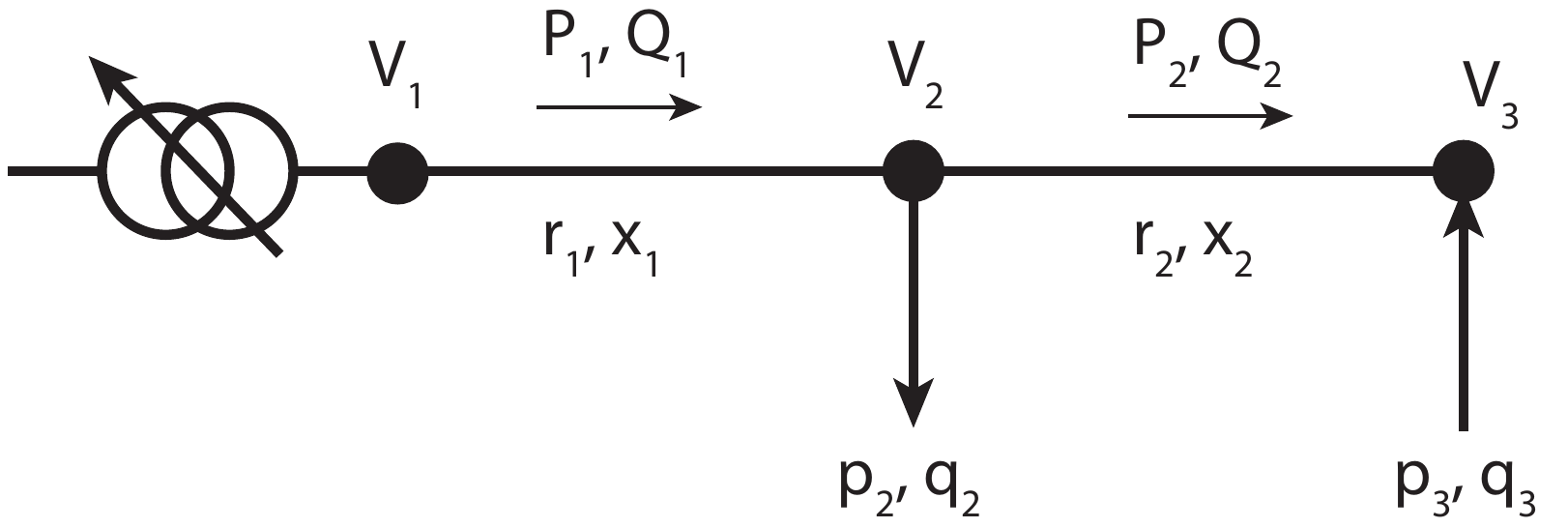}
\caption{The simplified distribution feeder line used in this study. Node 1 is at the substation where the SCADA enables control over $V_1$ via a tap changer. Node 2 represents a large aggregate real $p_2$ and reactive $q_2$ loads that fluctuate within a narrow range. Node three represents a distributed generator with real and reactive outputs $p_3$ and $q_3$. The assume the SCADA system enables control over $q_3$ to assist with voltage regulation along the circuit and that the attacker has taken control over $q_3$. The distribution circuit segments between the nodes have resistance $r_i$ and reactance $x_i$. The node injections $p_i$ and $q_i$ contribute to the circuit segment line flows $Q_i$ and $P_i$. }
\label{fig:ep}
\end{figure}

To keep the focus of this work on modeling the adversarial interaction between the defender and attacker, we retain the simplified model of an electrical grid used in previous work\cite{NIPS_Chapter}. Specifically, we consider the three-node model of a radial distribution circuit shown schematically in Fig.~\ref{fig:ep}.
The circuit starts at the  under-load tap changer (ULTC) on the low-voltage side of a substation transformer at node 1, serves an aggregation of loads at node 2, and connects a relatively large, individually-modeled distributed generator at node 3. In practice, most systems are considerably more detailed than this example. This example was chosen to limit the degrees of freedom to allow full enumeration of the parameter space and improve our understanding of the model's salient features. However, it is important to note that the model is not limited computationally by the size of the power system, rather it is limited by the number of players and their possible observations and actions. Extensions to more complex settings is an open challenge for future research.

In Fig.~\ref{fig:ep}, $V_i,p_i,$ and  $q_i$ are the voltage and real and reactive power injections at node $i$. $P_i, Q_i, r_i,$ and $x_i$ are the real power flow, reactive power flow, resistance, and reactance of circuit segment $i$. For this simple setting, we use the {\it LinDistFlow} equations~\cite{89BWa}
\begin{eqnarray}
\label{eq:powerEquation1}
P_2=-p_3,\;\; Q_2=-q_3,\;\;P_1=P_2+p_2,\;\;Q_1=Q_2+q_2\\
\label{eq:powerEquation2}
V_2=V_1-(r_1 P_1 + x_1 Q_1),\;\;V_3=V_2-(r_2 P_2 + x_2 Q_2).
\end{eqnarray}
Here, all terms have been normalized by the nominal system voltage $V_0$, and we set $r_i=0.03$ and $x_i=0.03$.

The attacker-defender game is modeled in discrete time with each simulation step representing one minute. To emulate the normal fluctuations of consumer real load, $p_2$ at each time step is drawn from a uniform distribution over the relatively narrow range $[p_{2,min},p_{2,max}]$ with $q_2=0.5 p_2$. The real power injection $p_3$ is of the distributed generator at node 3 is fixed. Although fixed for any one instance of the game, $p_{2,max}$ and $p_3$ are our design parameters, and we vary these parameters to study how they affect the outcome of the attacker-defender game. In all scenarios, $p_{2,min}$ is set 0.05 below $p_{2,max}$

In our simplified game, the SCADA operator (defender) tries to, keep the voltages $V_2$ and $V_3$ within appropriate operating bounds (described in more detail below). Normally, the operator has two controls: the ULTC to adjust the voltage $V_1$ or the reactive power output $q_3$ of the distributed generator. We assume that the system has been compromised, and the attacker has control of $q_3$ while the defender retains control of $V_1$. Changes in $V_1$ comprise the defender decision node while control of $q_3$ comprise the attacker decision node.

By controlling $q_3$, the attacker can modify the $Q_i$ and cause the voltage $V_2$ at the customer node to deviate significantly from $1.0\;p.u.$ -- potentially leading to economic losses by damaging customer equipment or by disrupting computers or computer-based controllers belonging to commercial or industrial customers\cite{LBNL-PQ}. The attacker's goals are modeled by the reward function
\begin{equation}
\label{eq:attackerReward}
R_A= \Theta (V_2-(1+\epsilon)) + \Theta ((1-\epsilon)-V_2).
\end{equation}
Here, $\epsilon$ represents the halfwidth of the acceptable range of normalized voltage. For most distribution systems under consideration, $\epsilon \sim 0.05$. $\Theta(\cdot)$ is a step function representing the need for the attacker to cross a voltage deviation threshold to cause damage.

In contrast, the defender attempts to control both $V_2$ and $V_3$ to near $1.0\;p.u.$. The defender may also respond to relatively small voltage deviations that provide no benefit to the attacker. We express these defender goals through the reward function
\begin{equation}
\label{eq:defenderReward}
R_D=-\left (\frac{V_2-1}{\epsilon}\right)^2  -\left (\frac{V_3-1}{\epsilon}\right)^2.
\end{equation}

\section{Time-Extended, Iterated Semi Net-Form Game}

To predict how system design choices affect the outcome of attacker-defender interactions, we need a description of when player decisions are made and how these decisions affect the system state, i.e. a ``game'' definition. Sophisticated attacker strategies may be carried out over many time steps (i.e. many sequential decisions), therefore we need to expand the SNFG description in the Introduction to allow for this possibility.

\begin{figure}
\begin{center}
\includegraphics[scale=.50]{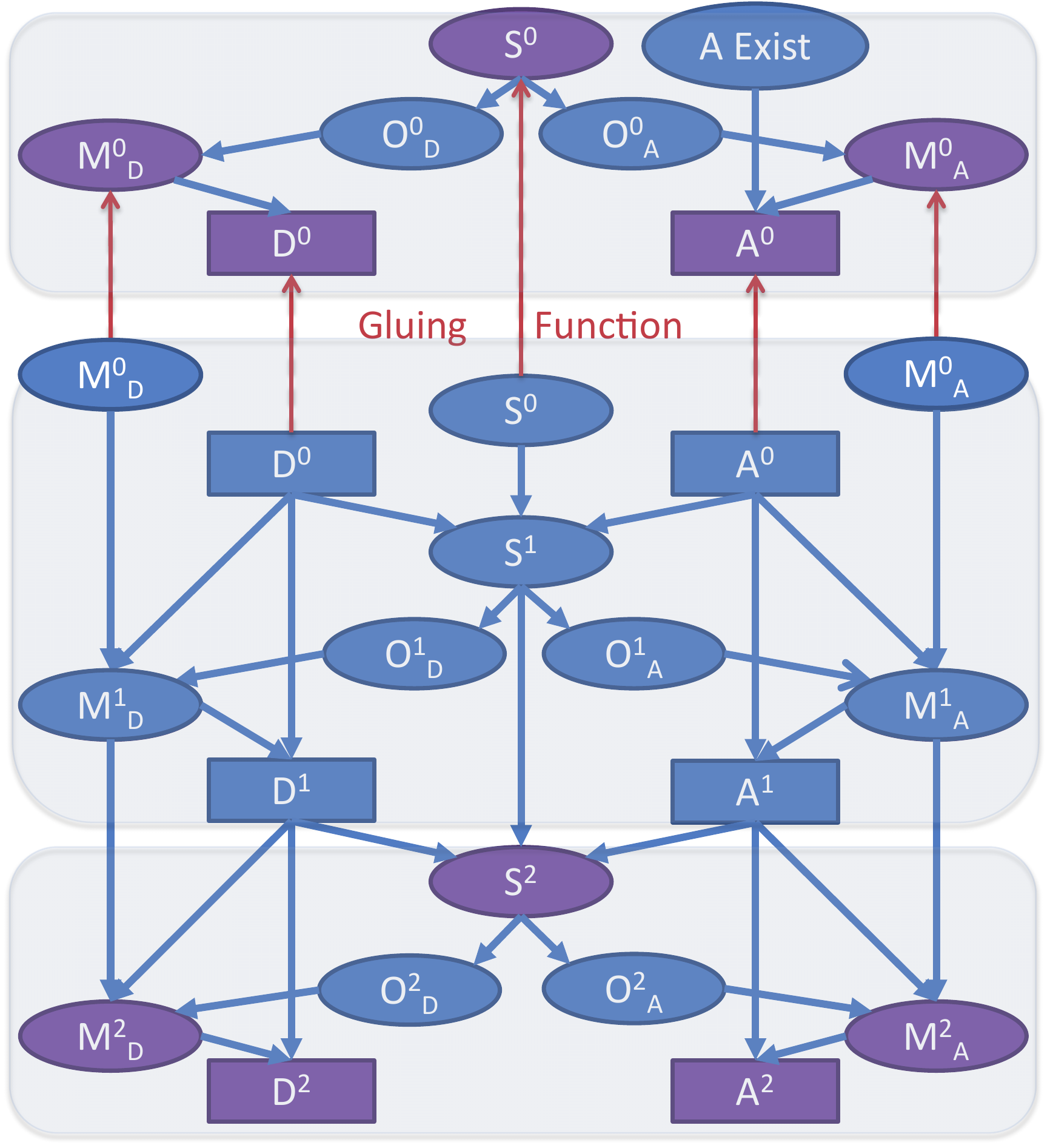}
\end{center}
\caption{The iterated semi net-form game (SNFG) used to model attackers and operators/defenders in a cyber-physical system. The iterated SNFG in the Figure consists of three individual SNFGs that are ``glued'' together at a subset of the nodes in the semi Bayes net that make up each SNFG. }
\label{fig:bayes}
\vspace{-0.25 in}
\end{figure}

Figure~\ref{fig:bayes} shows three individual semi-Bayes networks representing three time steps of our time-extended attacker-defender interaction. Each semi-Bayes net has the structure of a distinct SNFG played out at time step $i$. These SNFGs are ``glued'' together to form an iterated SNFG by passing the system state $S^i$, the players' moves/decisions $D_D^i$ and $D_A^i$, and the players' memories $M_D^i$ and $M_A^i$ from the SNFG at time step $i$ to the SNFG at time step $i+1$. Iterated SNFGs are described in more detail in \cite{NIPS_Chapter}.

In the rest of this section, we describe the nodes in this iterated SNFG and their relationship to one another.

\subsubsection{Attacker existence}
In contrast to our previous work, we add an `A exist'' node in Fig.~\ref{fig:bayes}--the only node that is not repeated in each SNFG.  This node contains a known probability distribution that outputs a $1$ (attacker exists) with probability $p$ and a $0$ (no attacker) with probability $1-p$.  When the attacker is not present, his decision nodes ($D_A^i$) are disabled and $q_3$ is not changed. We vary $p$ to explore the effect of different attack probabilities.

\subsubsection{System state}
The nodes $S^i$ contain the true physical state of the cyber-physical system at the beginning of the time step $i$. We note that the defender's memory $M_D^i$ and the attacker's memory $M_A^i$ are explicitly held separate from the $S^i$ to indicate that they cannot observed by other player.

\subsubsection{Observation Spaces}
Extending from $S^i$ are two directed edges to defender and attacker observation nodes $O^i_D$ and $O^i_A$.  The defender and attacker observation spaces, $\Omega_D$ and $\Omega_A$, respectively, are
\begin{equation}
    \Omega_D=[V_1,V_2,V_3,P_1,Q_1], \;\; \Omega_A= [V_2,V_3,p_3,q_3].
\end{equation}
These observations are not complete (the players do not get full state information), they may be binned (indicating only the range of a variable, not the precise value), and they may be noisy. The content of $\Omega_D$ and $\Omega_A$ is an assumption about the capabilities of the players. Here, $\Omega_D$ provides a large amount of system visibility consistent with the defender being the SCADA operator. However, it does not include $p_3$ or $q_3$ as the distributed generator has been taken over by the attacker. In contrast, $\Omega_A$ mostly provides information about node 3 and also includes $V_2$ because a sophisticated attacker would be able to estimate $V_2$ from the other information in $\Omega_A$. Although we do not consider this possibility here, we note that the content of the $\Omega_D$ and to some extent the content of $\Omega_A$ are potential control system design variables that would affect the outcome of the attacker-defender interaction. For example, excluding $V_3$ from $\Omega_D$ will affect the decisions made by the defender, and therefore, the outcome of the interaction.

\subsubsection{Player Memories}
The content and evolution of player memories should be constructed based on application-specific domain knowledge or guided by human-based experiments. In this initial work, we assume a defender memory $M_D^i$ and attacker memory $M_A^i$  consisting of a few main components
\begin{equation}
M^i_D=[\Omega^i_D,D^{i-1}_D,\mathcal{M}^i_D]; \;\;M^i_A=[\Omega^i_A,D^{i-1}_A,\mathcal{M}^i_A].
\end{equation}
The inclusion of the player's current observations $\Omega^i$ and previous move $D^{i-1}$ are indicated by directed edges in Fig.~\ref{fig:bayes}. The directed edge from $M^{i-1}$ to $M^i$ indicates the carrying forward and updating of a summary metric $\mathcal{M}^i$ that potentially provides a player with crucial additional, yet imperfect, system information that cannot be directly observed.

Our defender uses $\mathcal{M}_D$ to estimate if an attacker is present. One mathematical construct that provides this is
\begin{eqnarray}
\mathcal{M}^i_{D}&=& (1-1/n) \mathcal{M}^{i-1}_{D}\nonumber\\
&+& \textrm{sign}(V^i_{1}-V^{i-1}_{1})\;\textrm{sign}(V^i_{3}-V^{i-1}_{3}) \label{eq:defenderMemory}
\end{eqnarray}
The form of statistic in Eq.~\ref{eq:defenderMemory} is similar to the exponentially decaying memory proposed by Lehrer\cite{Lehrer88}. For attackers with small $q_3$ capability, even full range changes of $q_3$ will not greatly affect $V_3$, and the sign of changes in $V_3$ will be the same those of $V_1$.  The second term on the RHS of Eq.~\ref{eq:defenderMemory} will always be +1, and $\mathcal{M}_D\rightarrow 1$. An attacker with large $q_3$ capability can drive changes in $V_1$ and $V_3$ of opposite sign.  Several sequential time steps of with opposing voltage changes will cause $\mathcal{M}_D\rightarrow -1$. We note that if the defender does not change $V_1$, the contribution to $\mathcal{M}_D$ is zero, and the defender does not gain any information.

The general form of the attacker's memory statistic is similar to the defender's,
\begin{eqnarray}
\mathcal{M}^i_{A}&=& (1-1/n) \mathcal{M}^{i-1}_{A}\nonumber\\
&+& \textrm{sign}\left (\textrm{floor} \left ( \frac{\Delta V^i_{3}-\Delta q^i_{3} x_2/V_0}{\delta v} \right ) \right ),
\label{eq:attackerMemory}
\end{eqnarray}
however the contributions to $\mathcal{M}_A$ are designed to track the defender's changes to $V_1$. If the attacker changes $q_3$ by $\Delta q^i_{3}=q^i_{3}-q^{i-1}_{3}$, the attacker would expect a proportional change in $V_3$ by $\Delta V^i_{3}=V^i_{3}-V^{i-1}_{3}\sim \Delta q^i_3 x_2/V_0$. If $V_3$ changes according to this reasoning, then the second term on the RHS of Eq.~\ref{eq:attackerMemory} is zero. If instead the defender simultaneously increases $V_1$ by $\delta v$, $\Delta V^i_{3}$ will increase by $\delta v$, and the second term on the RHS of Eq.~\ref{eq:attackerMemory} is then +1. A similar argument yields -1 if the defender decreases $V_1$ by $\delta v$. Equation~\ref{eq:attackerMemory} then approximately tracks the aggregate changes in $V_1$ over the previous $n$ time steps.

\subsubsection{Decision or Move space}
Here, we only describe the decision options available to the players. {\it How} decisions are made is discussed in the next Section. Typical hardware-imposed limits of a ULTC constrain the defender actions at time step $i$ to the following domain
\begin{equation}
\label{eq:defenderDecision}
D^i_{D} = \{ \min(v_{max}, V^i_{1} + \delta v )   , V^i_{1},  \max(v_{min}, V^i_{1} - \delta v ) \}
\label{eq:defendermovespace}
\end{equation}
where $\delta v$ is the voltage step size for the transformer, and $v_{min}$ and $v_{max}$ represent the absolute min and max voltage the transformer can produce. In simple terms, the defender may leave $V_1$ unchanged or move it up or down by $\delta v$ as long as $V_1$ stays within the range $[v_{min},v_{max}]$. We take $v_{min}=0.90$, $v_{max}=1.10$, and $\delta v=0.02$. We allow a single tap change per time step (of one minute) which is a reasonable approximation tap changer lockout following a tap change.

Hardware limitations on the generator at node 3 constrain the attacker's range of control of $q_3$.  In reality, these limits can be complicated, however, we simplify the constraints by taking the attacker's $q_3$ control domain to be
\begin{equation}
\label{eq:attackerDecision}
D^i_{A} = \{- p_{3,max}, \ldots, 0, \ldots, p_{3,max}\}.
\end{equation}
In principle, the attacker could continuously adjust $q_3$ within this range. To reduce the complexity of our computations, we discretize the attacker's move space to eleven equally-spaced settings with $-p_{3,max}$ and $+p_{3,max}$ as the end points.

\section{Solution Concepts}

Nodes other than $D^i_D$ and $D^i_A$ represent control algorithms, evolution of a physical system, a mechanistic memory model, or other conditional probability distributions that can be written down without reference to any of the other nodes in the semi-Bayes net of Fig.~\ref{fig:bayes}. Specifying nodes $D^i_D$ and $D^i_A$ requires a model of human decision making. In an iterated SNFG with $N$ time steps, our defender would $3^N$ possibilities, and maximizing his average reward ($\sum_{i=1}^N R_D^i/N$) quickly becomes computationally challenging for reasonably large $N$. However, a human would not consider all $3^N$ choices.  Therefore, we seek a different solution concept that better represents human decision making, which is then necessarily tractable.

\subsection{Policies}

We consider a policy-based approach for players' decisions, i.e. a mapping from a player's memory to his action ($M^i_D\rightarrow D_D^i$). A single decision regarding what policy to use for the entire iterated SNFG greatly reduces the complexity making it independent of $N$.  A policy does not dictate the action at each time step.  Rather, the action at time step $i$ is determined by sampling from the policy based on the actual values of $M^i_D$ ($M^i_A$). We note that the reward garnered by a player's single policy decision depends on the policy decisions of other player because the reward functions of both players depend on variables affect by the other player's policy. Policies and the methods for finding optimal policies are discussed in greater detail in \cite{NIPS_Chapter}.

\subsection{Solution Concept: Level-K Reasoning}
The coupling between the players' policies again increases the complexity of computing the solution. However, the fully rational procedure of a player assessing his own reward based on all combinations of the two competing policies is not a good model of human decision making. We remove this coupling by invoking level-k reasoning as a solution concept. Starting at the lowest level-k, a level-1 defender policy is determined by finding the policy that maximizes the level-1 defender average reward when playing against a level-0 attacker. Similarly, the level-1 attacker policy is determined by optimizing against a level-0 defender policy. The higher k-level policies are determined by optimization with regard to the the k-1 policies. We note that the level-0 policies cannot be determined in this manner. They are simply assumptions about the non-strategic policy behavior of the attacker and defender that are inputs to this iterative process.

From the perspective of a level-k player, the decision node of his level k-1 opponent is now simply a predetermined conditional probability distribution making it no different than any other node in the iterated SNFG, i.e. simply part of his environment. The level-k player only needs to compute his best-response policy against this fixed level k-1 opponent/environment.  The selection of the levek-k policy is now a single-agent reinforcement learning problem. Level-k reasoning as a solution concept is discussed in more detail in \cite{NIPS_Chapter}.

\subsection{Reinforcement Learning}

Many standard reinforcement learning techniques can be used to solve the optimization problem discussed above~\cite{Busoniu10,KaelblingSurvey,SuttonBarto98}. In our previous work\cite{NIPS_Chapter}, the attacker and defender optimization problems were both modeled as Markov Decision Processes (MDP), even though neither player could observe the entire state of the grid. The additional uncertainty related to attacker existence casts doubt on this approach. Instead, we employ a reinforcement learning algorithm based on \cite{Jaakkola} which has convergence guarantees for Partially Observable MDPs (POMDP). This approach has two distinct steps. First is the policy evaluation step, where the Q-values for the current policy are estimated using Monte Carlo. Second, the policy is updated by placing greater weight on actions with higher estimated Q-values. The two steps are iterated until the policy converges to a fixed point indicating a local maximum has been found. The details of the algorithm can be found in \cite{Jaakkola}.

\section{Simulation Results}

Due to space limitations and our desire to explore the design aspects of our models, we only consider results for a level-1 defender matched against a level-0 attacker. We retain the level-0 attacker policy that we have used in our previous work\cite{NIPS_Chapter}. Although he is only level-0, this level-0 attacker is modeled as being knowledgeable about power systems and is sophisticated in his attack policy.

\subsection{Level-0 Attacker}
The level-0 attacker drifts one step at at time to larger $q_3$ if $V_2<1$ and smaller $q_3$ if $V_2>1$. The choice of $V_2$ to decide the direction of the drift is somewhat arbitrary, however, this is simply assumed level-0 attacker behavior. The drift in $q_3$ causes a drift in $Q_1$ and, without any compensating move by the defender, a drift in $V_2$.  A level-1 defender that is unaware of the attacker's presence would compensate by adjusting $V_1$ in the opposite sense as $V_2$ in order to keep  the average of $V_2$ and $V_3$ close to 1.0. The level-0 attacker continues this slow drift forcing the unaware level-1 defender to ratchet $V_1$ near to $v_{min}$ or $v_{max}$. At some point, based on his knowledge of the power flow equations and the physical circuit, the level-0 attacker determines it is time to ``strike'', i.e. a sudden large change of $q_3$ in the opposite direction to the drift would push $V_2$ outside the range $[1-\varepsilon,1+\epsilon]$. If the deviation of $V_2$ is large, it will take the defender a number of time steps to bring $V_2$ back in range, and the attacker accumulates reward during this recovery time. More formally, this level-0 attacker policy can be expressed as
\begin{algorithm}{Level0Attacker}{}
 V^* = \max_{q \in D_{A,t}} |V_2-1|;\\
\begin{IF}{V^* > \theta_A}
  \RETURN \arg\max_{q \in D_{A,t}} |V_2-1|;
\end{IF}\\
\begin{IF}{V_2 < 1}
  \RETURN q_{3,t-1}+1;
\end{IF}\\
\RETURN q_{3,t-1}-1;
\end{algorithm}
Here, $\theta_A$ is a threshold parameter that triggers the strike. Throughout this work, we have used $\theta_A=0.07> \epsilon $ to indicate when an attacker strike will accumulate reward.

\subsection{Level-1 Defender--Level-0 Attacker Dynamics}

\begin{figure}
\begin{center}
\includegraphics[scale=.37]{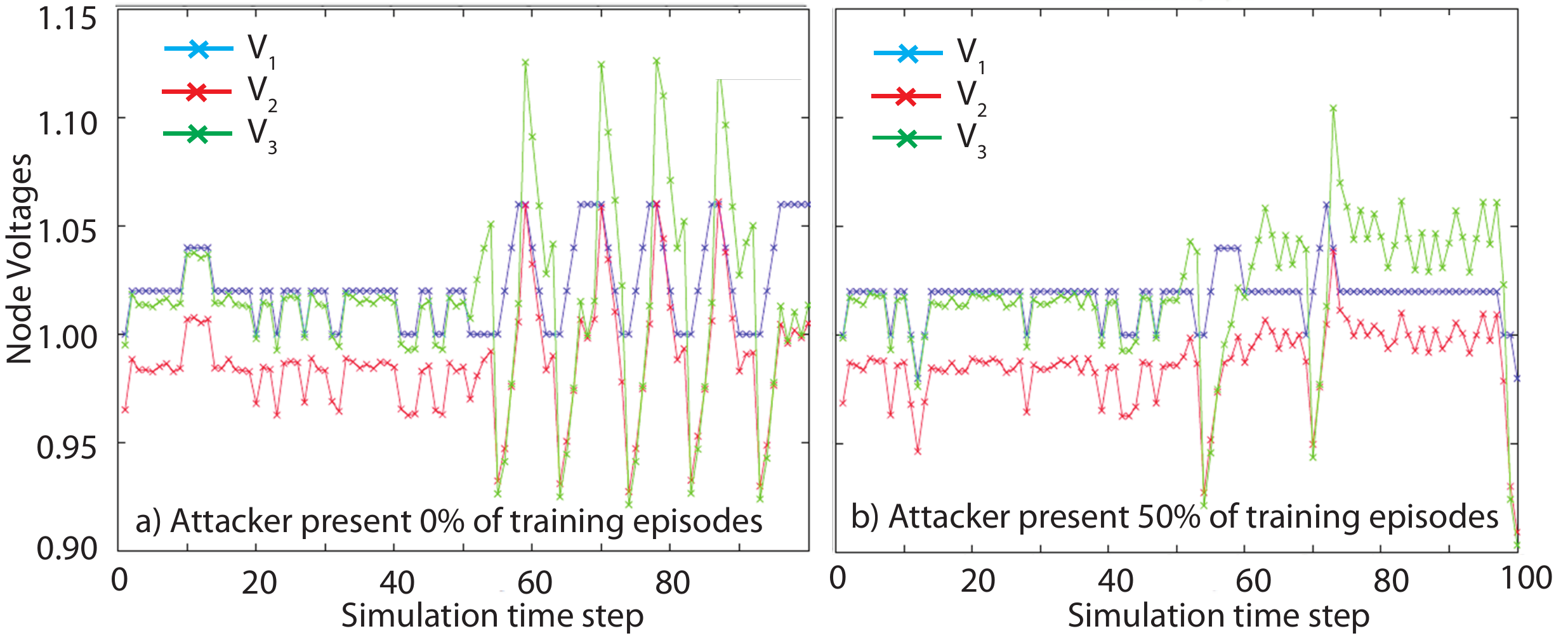}
\end{center}
\caption{Typical time evolution of $V_1$ (blue), $V_2$ (red), and $V_3$ (green) for a level-1 defender facing a level-0 attacker. In the left plot, the level-1 defender's policy was optimized for $p=0.0$ in ``A exist'', i.e. no level-0 attacker was ever present. In the right plot, the level-1 defender's policy was optimized with $p=0.50$. At the start of the simulation, no attacker is present. The attacker enters the simulation at time step 50. In these simulations, $p_{2,max}=1.4$, $p_{2,min}=1.35$ and $p_{3,max}=1.0$.
}
\label{fig:voltage_results}
\end{figure}

We demonstrate our entire modeling and simulation process on two cases. In the first case, a level-1 defender optimizes his policy against a level-0 attacker that is present 50\% of the time, i.e. $p=0.50$ in the node ``A exist'' in Fig.~\ref{fig:bayes}. In the second case, the level-1 defender optimizes his policy against a ``normal'' system, i.e. $p=0.0$ in ``A exist''. The behavior of these two level-1 defenders is shown in Fig.~\ref{fig:voltage_results} where we temporarily depart from the description of our model. In the first half of these simulations, the level-0 attacker does not exist, i.e. $p=0.0$, and there are no significant differences between the two level-1 defenders. At time step 50, a level-0 attacker is introduced with $p=1.0$. The level-1 defender optimized for $p=0.0$ suffers from the ``drift-and-strike'' attacks as described above.  In contrast, the level-1 defender with a policy optimized at $p=0.50$ has learned not to follow these slow drifts and maintains a more or less steady $V_1$ even after time step 50. Although $V_3$ is out of acceptable bounds for some periods, these are much shorter than before and $V_2$ is never out of bounds.

\subsection{Policy Dependence on $p$ During Defender Training}
Next, we present a few preliminary studies that prepare our model for studying circuit design tradeoffs. Although policy optimization (i.e. training) and policy evaluation seem closely related, we carry these out as two distinct processes. During training, all of the parameters of the system are fixed, especially the probability of the attacker presence $p$ and the circuit parameters $p_{2,max}$ and $p_{3,max}$. Many training runs are carried out and the policy is evolved until the reward per time step generated by the policy converges to a fixed point. The converged policy can then be evaluated against the conditions for which is was trained, and in addition, it can be evaluated for different but related conditions. For example, we can train with one probability of attacker existence $p$, but evaluate the policy against a different value of $p^\prime$. Next, we carry out just such a study to determine if a single value of $p$ can used in all of our subsequent level-1 defender training. If using a single training $p$ can be justified, it will greatly reduce the parameter space to explore during subsequent design studies.

\begin{figure}
\begin{center}
\includegraphics[scale=.285]{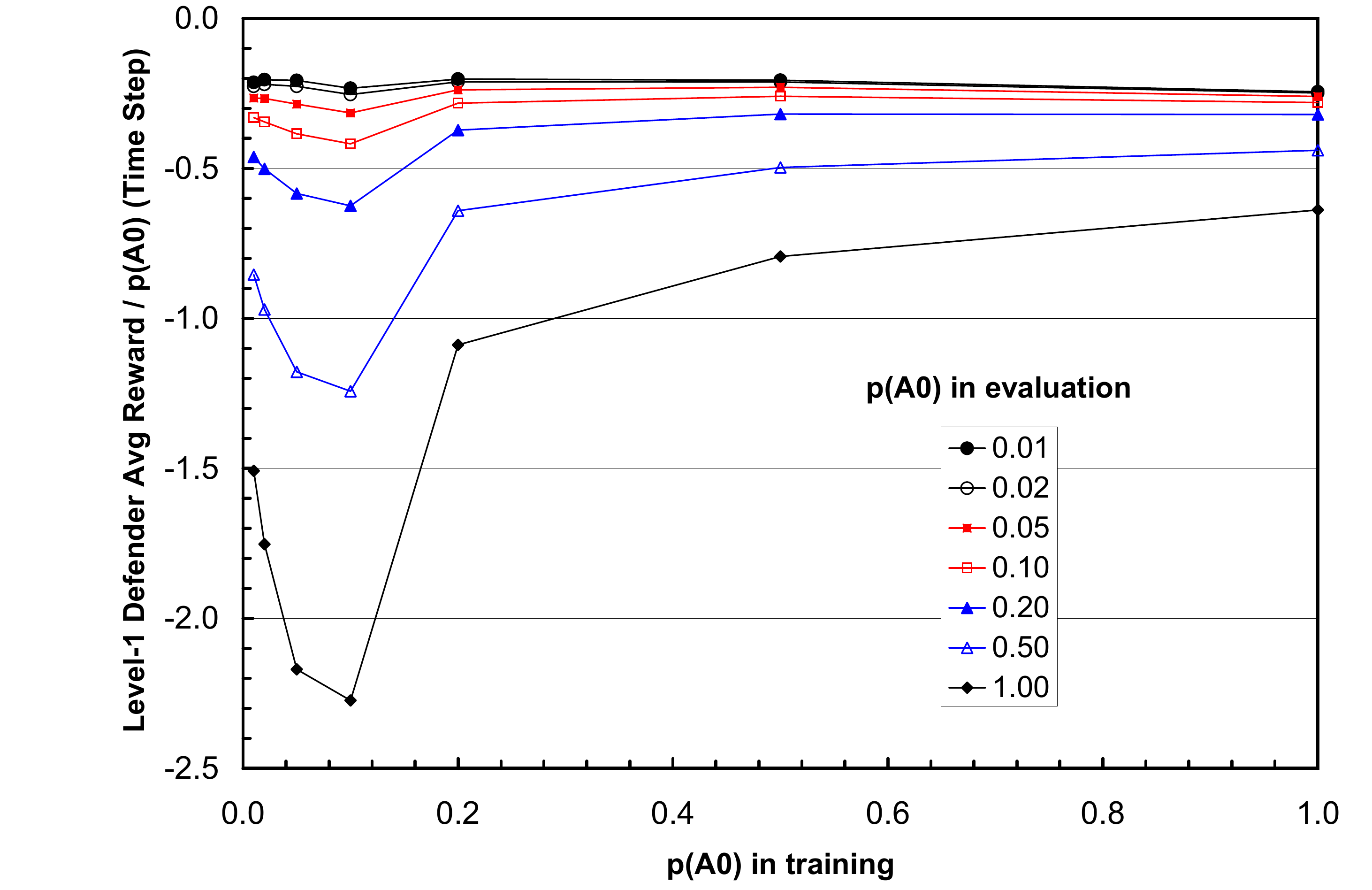}
\end{center}
\caption{Level-1 defender reward per time step of level-0 attacker presence during {\it simulation} ($\sum_{i=1}^N R_D^i/Np$ ) versus the probability of attacker presence during {\it training}. The curves representing different levels of attacker presence during {\it simulation} all show the same general dependence, i.e. a relatively flat plateau in normalized level-1 defender reward for $p\geq 0.20$ due to the more complete sampling of the states $S$ during training (i.e. policy optimization). This common feature leads us to select $p=0.20$ for training for all the subsequent work in the manuscript.
}
\label{fig:p(a0)study}
\end{figure}

We consider seven values of $p$ logarithmically spaced from 0.01 to 1.0. A set of seven level-1 defenders, one for each $p$, is created by optimizing their individual policies against a level-0 attacker who is present with probability $p$. Each of these defenders is then simulated seven times, i.e. against the same  level-0 attacker using the same range of $p$ as in the training. In these simulations, $p_{2,max}=1.4$ and $p_{3,max}=1.0$. During the simulation stage, the average defender reward per time step is computed and normalized by the value of $p$ during the simulation stage, i.e. $\sum_{i=1}^N R_D^i/Np$, creating a measure of level-1 defender performance per time step that the level-0 attacker is actually present. The results are shown in Fig.~\ref{fig:p(a0)study}.

For an achievable number of Monte Carlo samples and for low values of $p$ during policy optimization (i.e. training), there will be many system states $S$ that are visited infrequently or not at all, particularly those states where the attacker is present. The reinforcement learning algorithm will provide poor estimates of the Q values for these states, and the results of the policy optimization should not be trusted. For these infrequently visited states, we replace the state-action policy mapping with the mapping given by the level-0 defender policy used in our previous work\cite{NIPS_Chapter}. Even with this replacement, the level-1 defenders trained with $p<0.10$ perform quite poorly. For $p\geq 0.20$, it appears that enough states are visited frequently enough such that level-1 defender performance improves. For the remainder of the studies in this manuscript, we use $p=0.20$ for all of our level-1 defender training.

\section{Design Procedure and Social Welfare}

Significant deviations of $V_2$ or $V_3$ from 1.0 p.u. can cause economic loss either from equipment damage or lost productivity due to disturbances to computers or computer-based industrial controllers\cite{LBNL-PQ}. The likelihood of such voltage deviations is increased because possibility of attacks on the distributed generator at node 3. However, this generator also provides a social benefit through the value of the energy it contributes to the grid. The larger the generator (larger $p_{3,max}$) the more energy it contributes and the higher this contribution to the social welfare. However, when compromised, a larger generator increases the likelihood of large voltage deviations and significant economic loss.

To balance the value of the energy against the lost productivity, we assess both in terms of dollars. The social welfare of the energy is relatively easy to estimate because the value of electrical energy, although variable in both time and grid location, can be assigned a relatively accurate average value. Here the value of electrical energy is approximated by a flat-rate consumer price. In heavily regulated markets, the price of electricity can be distorted, and this approach may be a bad approximation of the true value of the energy. So while price is a reasonable approximation of value for the purposes of this model, in practice it may be necessary to adjust for market distortions on a case-by-case basis. In our work, the generator at node 3 is installed in a distribution system where we estimate the energy value at $C_E=$\$80/MW-hr.

As with estimating the value of energy, estimating the social cost of poor power quality is also a prerequisite to the power grid design procedure. In contrast to the value of energy, there is no obvious proxy for this cost making it difficult to estimate.  Studies\cite{LBNL-PQ} have concluded that the cost is typically dominated by a few highly sensitive customers, making this cost also dependent on grid location and time -- the location of the highly sensitive customers and their periods operation drive this variability. The average cost of a power quality event has been estimated\cite{LBNL-PQ} at roughly $C_{PQ}=$\$300/sensitive customer/per power quality event. Note that social welfare, including estimates of the value of energy and the social cost of poor power quality, determines the optimality of the power grid design and should be carefully chosen for each application.

We now describe a series of numerical simulations and analyses that enable us to find the social welfare break even conditions for the generator at node 3.

\subsection{Level-1 Defender Performance Versus ($p_{2,max}$,$p_{3,max}$)}

Because the output of the  node ``A exist'' in the iterated SNFG in Fig.~\ref{fig:bayes} fixes the probability of the presence of an attacker for the rest of the $N$ steps in the simulation, the results from each simulation of the iterated SNFG are statistically independent. Therefore, if we know the level-1 defender's average reward when he is under attack 100\% of the time ($p=1$) and 0\% of the time ($p=0$), we can compute his average reward for any intermediate value of $p$. Taking this into account, we proceed as follows. Using the guidance from the results in Fig.~\ref{fig:p(a0)study}, we train level-1 defenders against level-0 attackers (using $p=0.2$) for an array of ($p_{2,max}$,$p_{3,max}$) conditions. Next, we simulate these level-1 defenders with $p=1$ and $p=0$ so that we can compute their average reward for any $p$. The results for all ($p_{2,max}$,$p_{3,max}$) conditions for $p=0.01$ are shown in Fig.~\ref{fig:RD_vs_p3}. The results show two important thresholds, i.e. the level-1 defenders' average reward falls off quickly when $p_{3,max}>1.5$ or when $p_{2,max}>1.9$. In the rest of this analysis, we will focus on the region $p_{3,max}<1.5$ before the large decrease in the defender's average reward

\subsection{Level-1 Defender $p_{3,max}$ Sensitivity }

\begin{figure}
\vspace{-.25cm}
\begin{center}
\includegraphics[scale=.285]{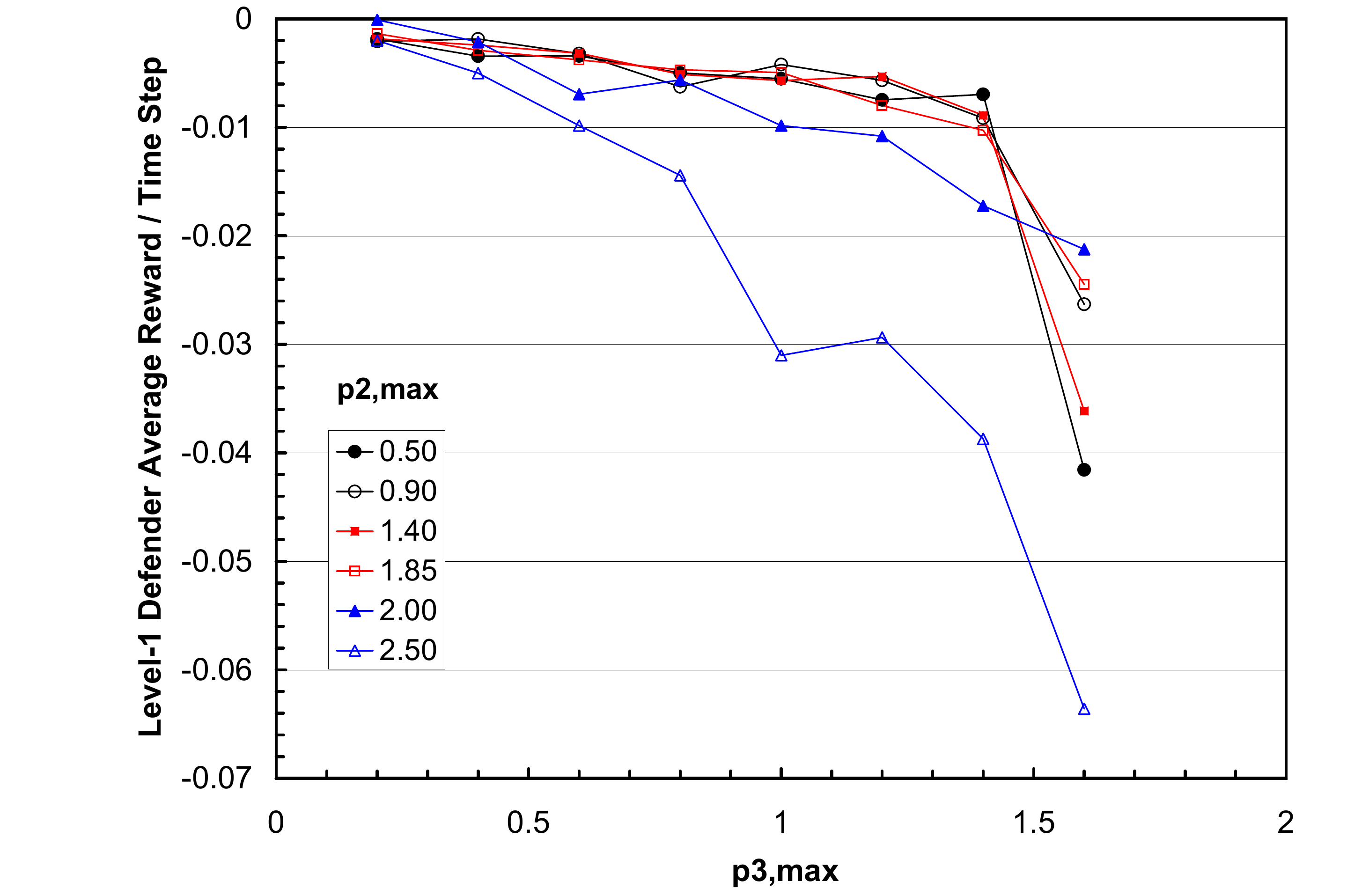}
\vspace{-.25cm}
\end{center}
\caption{The level-1 defender's average reward per simulation time step as a function of $p_{3,max}$ for a 1\% probability of an attack on node 3. Each curve represents a different value of $p_{2,max}$ in the range $[0.2...2.5]$. 
}
\label{fig:RD_vs_p3}
\end{figure}

\begin{figure}
\vspace{-.5cm}
\begin{center}
\includegraphics[scale=.285]{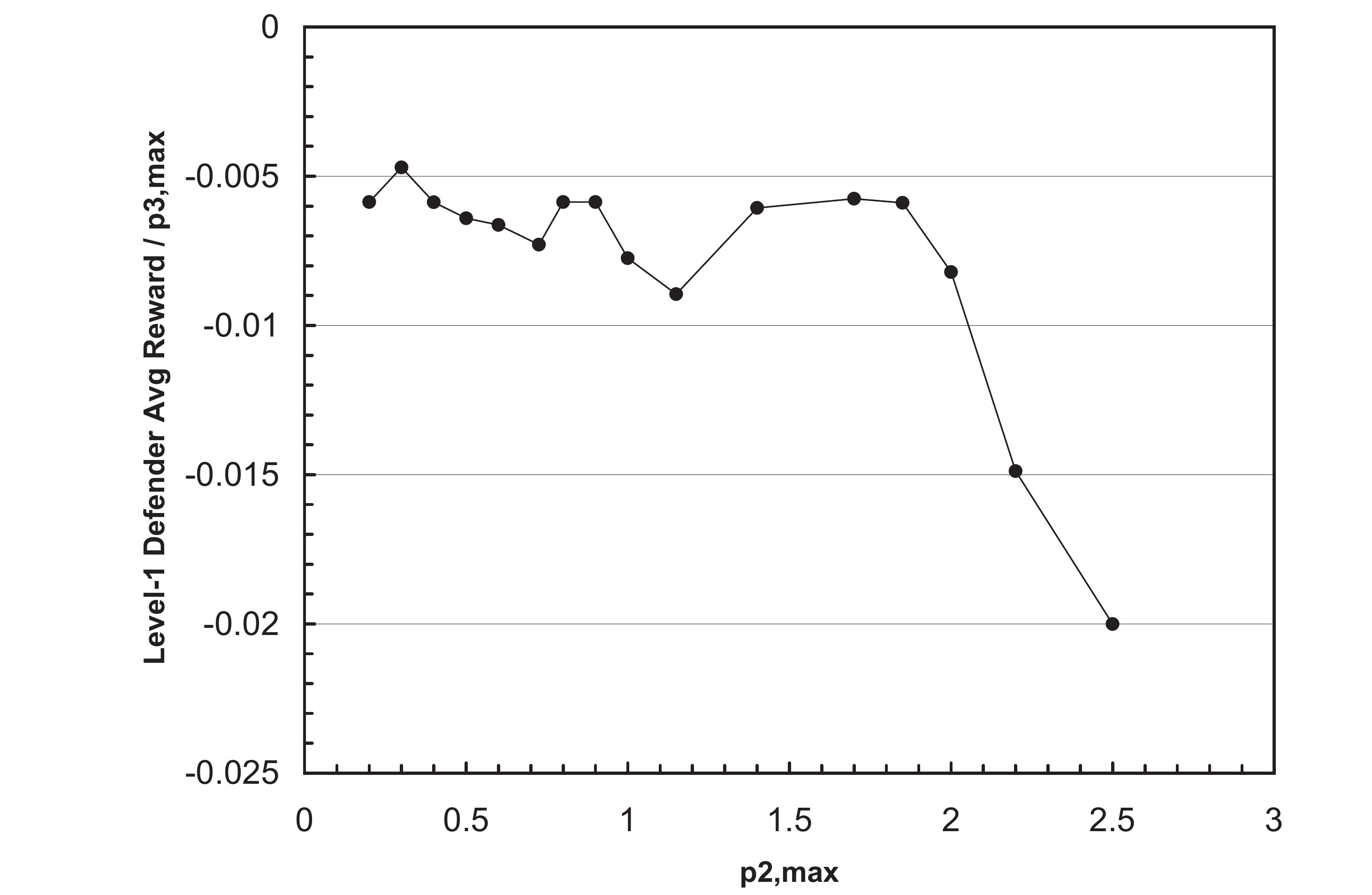}
\vspace{-.5 cm}
\end{center}
\caption{The slope of the data in Fig.~\ref{fig:RD_vs_p3} for $p_{3,max}\leq 1.5$. The slope measures how quickly the level-1 defender's reward decreases with $p_{3,max}$ for different values of $p_{2,max}$. Consistent with Fig.~\ref{fig:RD_vs_p3}, the slope is roughly constant for $p_{2,max}<1.9$ and then rapidly becomes more negative as $p_{2,max}$ increases beyond 1.9. The rapid decrease demonstrates the level-1 defender is much more susceptable to the level-0 attacker when $p_{2,max}>1.9$.
}
\label{fig:RD_slope_vs_p2}
\end{figure}

Using the energy and power quality cost estimates from above, the results in Fig.~\ref{fig:RD_vs_p3} could be turned into surface plots of social welfare. However, the number of design parameters that could be varied would generate a multi-dimensional set of such surface plots making the results difficult to interpret. Instead, we seek to reduce this dimensionality and generate results that provide more design intuition. We first note that the level-1 defenders' reward falls approximately linearly with $p_{3,max}$ for $p_{3,max}<1.5$. The slope of these curves is the sensitivity of the level-1 defenders' average reward to $p_{3,max}$, and we extract and plot these sensitivities versus $p_{2,max}$ in Fig.~\ref{fig:RD_slope_vs_p2}.

To further analyze the results in Fig.~\ref{fig:RD_slope_vs_p2}, we must relate the defender's average reward to power quality events, which can then be converted into a social welfare cost using $C_{PQ}$. Equation~\ref{eq:defenderReward} expresses the defender's reward $R_D$ as a sum of two smooth functions (one function of $V_2$ and another of $V_3$). These individual contributions are equal to $1$ when $V_2$ or $V_3$ are equal to either $1+\epsilon$ or $1-\epsilon$. Although these deviations are not severe, we consider such deviations to constitute a power quality event, and we estimate its social welfare cost by as $R_D C_{PQ}$. $R_D$ increases (decreases) quadratically for larger (smaller) voltage deviations, and our definition of the social welfare cost captures that these larger (smaller) deviations result in higher (lower) social welfare costs. Using $C_{PQ}=$\$300/sensitive customer/per power quality event estimated in \cite{LBNL-PQ}, our simulation time step of one minute, and assuming there is one sensitive customer on our circuit, the slopes of $\sim$0.006/(MW of $p_{3,max}$) in Fig.~\ref{fig:RD_slope_vs_p2} corresponds to a social welfare cost of \$108/(MW of $p_{3,max})$/hr. At this value of $C_{PG}$, the social value provided by the energy at \$80/MW-hr is outweighed by the social welfare cost caused by the reduction in power quality.

\begin{figure}
\vspace{-.25cm}
\begin{center}
\includegraphics[scale=.285]{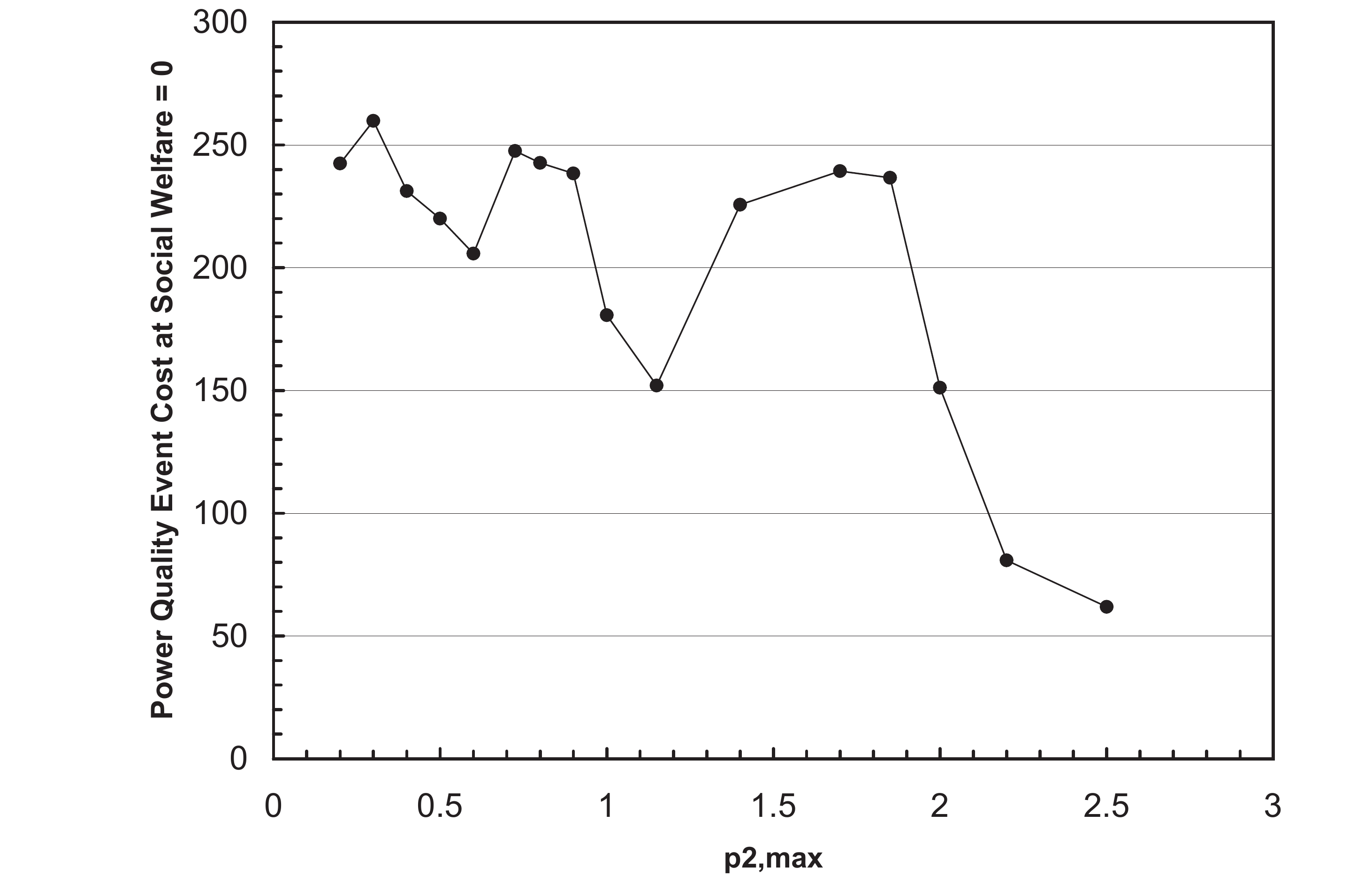}
\vspace{-.5 cm}
\end{center}
\caption{The cost of a power quality event that yields a zero social welfare contribution from the distributed generator at node 3 (i.e. the generator's break even point) versus $p_{2,max}$. To compute the break even cost of power quality events, we have assumed: the generator is under attack by a level-0 attacker 1\% of the time and the value of the energy from the generator at node 3 is \$80/MW-hr.
}
\label{fig:PQ_SWzero}
\end{figure}

Slightly modifying the analysis just described, we can now find the energy/power-quality break even points for the social welfare of the generator at node 3, i.e. the cost of a power quality event that reduces the social welfare provided by the energy to a net of zero. The break-even power quality cost is plotted versus $p_{2,max}$ in Fig.~\ref{fig:PQ_SWzero}. Points of $C_{PQ}$ and $p_{2,max}$ that fall to the lower left of the curve contribute positive social welfare while those to the upper right contribute negative social welfare. When applied to more realistic power system models, analysis such as shown in Fig.~\ref{fig:PQ_SWzero} can be used to make decisions about whether new distributed generation should be placed on a particular part of a distribution grid.

\section{Conclusion}

We have described a novel time-extended, game theoretic model of humans interacting with one another via a cyber-physical system, i.e. an interaction between a cyber intruder and an operator of an electrical grid SCADA system. The model is used to estimate the outcome of this adversarial interaction, and subsequent analysis is used to estimate the social welfare of these outcomes. The modeled interaction has several interesting features. First, the interaction is asymmetric because the SCADA operator is never completely certain of the presence of the attacker, but instead uses a simple statistical representation of memory to attempt to infer the attacker's existence. Second, the interaction is mediated by a significant amount of automation, and using the results of our model or related models, this automation can be (re)designed to improve the social welfare of these outcomes.

The models in this manuscript can be extended and improved in many ways. Perhaps the most important of these would be extending the model to incorporate larger, more realistic grids, such as transmission grids, where the meshed nature of the physical system would result in more complex impacts from an attack. In contrast to the setting described here, such complex grids would have multiple points where a cyber intruder could launch an attack, and models of the defender, his reward function, and his memory would be equally more complex.

As discussed earlier, one challenge with our approach is computational. The size of the physical system itself does not overly increase the computational requirements (beyond what is normally seen in solving power flow equations in large-scale systems).  However, the number of players and observations does increase the computational requirements exponentially. This is a major focus of our current work. In particular, we note that the number of observations (monitors) that a real human can pay attention to is very limited. One approach we are investigating for how to overcome the exponential explosion is to incorporate this aspect of real human limitations
into our model The challenge here will be developing models of how a human chooses which observations to make to guide their decisions.




\bibliographystyle{IEEEtran}
\bibliography{cyber,SmartGrid,ThesisBib,RLBib}

\end{document}